\documentclass{article}

\usepackage{arxiv}

\usepackage[utf8]{inputenc} 
\usepackage[T1]{fontenc}    
\usepackage{hyperref}       
\usepackage{url}            
\usepackage{booktabs}       
\usepackage{amsfonts}       
\usepackage{nicefrac}       
\usepackage{microtype}      
\usepackage{lipsum}
\usepackage{graphicx}

\usepackage[]{hyperref}
\usepackage{booktabs}
\usepackage{multirow}
\usepackage{makecell}

\graphicspath{ {./images/} }

\title{A Self-supervised Learning Method for Raman Spectroscopy based on Masked Autoencoders}

\author{
 Pengju Ren \\
  College of Information Engineering\\
  Shanghai Maritime University\\
  Shanghai, China 201306 \\
  \texttt{renpengju@stu.shmtu.edu.cn} \\
   \And
 Ri-gui Zhou \footnote{Corresponding author.} \\
  College of Information Engineering\\
  Shanghai Maritime University\\
  Shanghai, China 201306 \\
  \texttt{rgzhou@shmtu.edu.cn} \\
  \And
 Yaochong Li \\
  College of Information Engineering\\
  Shanghai Maritime University\\
  Shanghai, China 201306 \\
  \texttt{ycli@shmtu.edu.cn} \\
}

\begin{document}
\maketitle
\begin{abstract}
Raman spectroscopy serves as a powerful and reliable tool for analyzing the chemical information of substances. The integration of Raman spectroscopy with deep learning methods enables rapid qualitative and quantitative analysis of materials. Most existing approaches adopt supervised learning methods. Although supervised learning has achieved satisfactory accuracy in spectral analysis, it is still constrained by costly and limited well-annotated spectral datasets for training. When spectral annotation is challenging or the amount of annotated data is insufficient, the performance of supervised learning in spectral material identification declines. In order to address the challenge of feature extraction from unannotated spectra, we propose a self-supervised learning paradigm for Raman \textbf{S}pectroscopy based on a \textbf{M}asked \textbf{A}uto\textbf{E}ncoder, termed \textbf{SMAE}. SMAE does not require any spectral annotations during pre-training. By randomly masking and then reconstructing the spectral information, the model learns essential spectral features. The reconstructed spectra exhibit certain denoising properties, improving the signal-to-noise ratio (SNR) by more than twofold. Utilizing the network weights obtained from masked pre-training, SMAE achieves clustering accuracy of over 80\% for 30 classes of isolated bacteria in a pathogenic bacterial dataset, demonstrating significant improvements compared to classical unsupervised methods and other state-of-the-art deep clustering methods. After fine-tuning the network with a limited amount of annotated data, SMAE achieves an identification accuracy of 83.90\% on the test set, presenting competitive performance against the supervised ResNet (83.40\%). Unlike supervised learning, self-supervised learning does not rely on any data annotations and can learn more spectral features beyond annotation constraints, enabling the exploration of richer chemical information from substances. Self-supervised learning has the potential to become a powerful tool for spectral feature extraction and analysis.
\end{abstract}


\section{Introduction}
Raman spectroscopy is a molecular spectroscopic technique that provides in-depth insights into the composition and characteristics of materials through the interaction between light and matter. It provides benefits such as non-destructive analysis and label-free extraction of chemical information, making it highly valuable for the analysis of biological\cite{cheng_vibrational_2015,wang_multi-point_2023,kamp_raman_2024} and material\cite{guselnikova_pretreatment-free_2024,zhang_alkyne-tagged_2024} samples. Recently, Raman spectroscopy has emerged as a novel technology in clinical research and pathological bacterial analysis, demonstrating significant potential for rapid, label-free diagnosis during surgeries\cite{shin_single_2023,huang_rapid_2023,bresci_label-free_2024}. However, in practical applications, the Raman spectral signals are extremely weak and prone to overlap, it is difficult to distinguish with the naked eye. Thus, chemometric methods are typically employed for analysis\cite{xue_advances_2023}. In recent years, the rapid advancement of deep learning-based chemometrics has facilitated the application of Raman spectroscopy in the analysis of pathogenic bacteria, achieving high recognition accuracy in selecting appropriate antibiotic treatments\cite{lu_patch-based_2024,chen_combined_2024,deng_scale-adaptive_2022}. Most existing methods rely on traditional supervised learning approaches, which often involve complex neural network structures with parameters exceeding millions. This complexity can lead to data hunger, requiring a considerable amount of labeled data for training support. However, in medical pathological analysis, such as with liver samples, obtaining a large volume of spectral labels is particularly challenging, especially in dynamic systems where two or more consecutive experiments are needed to acquire training data pairs\cite{he_deep_2021}. In cancer case analysis, only a limited number of spectra are available for each patient’s tissue\cite{ma_classifying_2021}. Labelling spectra necessitates considerable time and effort from domain experts, and establishing a large, well-annotated spectral dataset is both time-consuming and costly. Furthermore, if the reference database changes, supervised learning networks require retraining, which increases computational costs\cite{liu_dynamic_2019}. Therefore, how to learn important features spaces from large amounts of unlabeled spectral data remains a persistent issue. The potential of unsupervised learning in spectral analysis has yet to be fully explored, and unsupervised learning is poised to become a powerful tool for processing spectral data.

When dealing with large amounts of unlabeled data, the first approach that comes to mind is classic unsupervised learning methods. In Raman spectroscopy, numerous researchers have proposed statistical methods such as principal components analysis (PCA) and t-distributed stochastic neighbor embedding (t-SNE) for dimensionality reduction and clustering of spectral data. These methods are often followed by K-means clustering to achieve classification of unlabeled spectral data\cite{sun_k-means_2020,arslan_discrimination_2022,uysal_ciloglu_identification_2020}. For instance, Sun et al.\cite{sun_k-means_2020} applied PCA for dimensionality reduction to extract the top three principal components from Raman spectra of extracellular vesicles, followed by K-means clustering of mouse mammary tissues and human breast cancer tissues, revealing spatial distribution and distinct chemical characteristics of extracellular vesicles. Similarly, Afra Hacer Arslan et al.\cite{arslan_discrimination_2022} visualized the three-dimensional principal component space of aquatic bacterial Raman spectra using PCA and employed hierarchical clustering analysis for rapid and unsupervised detection of bacterial pathogens. Fatma Uysal Ciloglu et al.\cite{uysal_ciloglu_identification_2020} also applied PCA for dimensionality reduction of pathogen bacterial Raman spectral features, followed by hierarchical clustering analysis to investigate methicillin-resistant bacteria. However, the optical system has the phenomenon of decreased efficiency in measurement, and the low efficiency of Raman scattering. As a result, Raman spectra of pathogenic bacteria often have a low signal-to-noise ratio (SNR)\cite{wang_applications_2021}. Moreover, the spectral data typically consist of thousands of variables, resulting in high-dimensional data, which makes traditional unsupervised methods more difficult to apply. Neural networks have demonstrated better performance in dimensionality reduction of high-dimensional spectral data compared to PCA and t-SNE. Guo et al.\cite{guo_unsupervised_2022} proposed a convolutional variational deep embedding clustering method (CVDE) to learn representation of Raman spectra and then employed gaussian mixture model (GMM) for soybean oil and drug Raman spectra clustering. Sun et al.\cite{sun_ramancluster_2024} introduced RamanCluster, a novel representation learning module, which utilized strong and weak spectral augmentations for a dual-level collaborative contrast learning approach to map high-dimensional data into low-dimensional space, and then applied an auxiliary deep embedded clustering component to classify pathogenic bacteria. However, as the amount of spectral data and the number of classifications increase, the accuracy of both methods falls short of expectations. Therefore, handling large data volumes, low SNR, and high-dimensional Raman spectral features poses significant challenges to traditional unsupervised learning methods.

Self-supervised learning is a subfield of unsupervised learning\cite{balestriero_cookbook_2023-1} that can circumvent the requirement for extensive labeled data in supervised learning, demonstrating broad application potential across various domains, such as images\cite{he_masked_2022,xie_simmim_2022,chen_context_2024}, videos\cite{tong_videomae_2022-1,feichtenhofer_masked_2022}, and speech\cite{huang_masked_2023,devlin_bert_2019}. Self-supervised learning constructs pretext tasks to create simulated data labels, thereby uncovering the intrinsic relationships between data points. In tandem mass spectrometry (MS/MS), Roman Bushuiev et al.\cite{bushuiev_emergence_2024} proposed a transformer-based neural network structure that employs self-supervised learning (DreaMS) on millions of unlabeled mass spectra. After fine-tuning the network, the DreaMS achieved predictions of spectral similarity, molecular fingerprints, chemical properties, and the presence of fluorine in tandem mass spectrometry. In Raman spectroscopy, Mathias N.Jensen et al.\cite{jensen_identification_2024} suggested introducing additional Gaussian noise, wavelength shifts, and clipping to disrupt the original spectra, training the network to recover the disrupted spectra to learn spectral features. However, this method has its drawbacks, as the strength of the added spectral noise and the magnitude of wavelength shifts are difficult to control, significantly affecting the performance of subsequent downstream classification tasks. Currently, research on self-supervised learning in Raman spectroscopy is still insufficient. When a large amount of labeled spectral data is not available, data augmentation is also a widely used method to address this issue.\cite{bjerrum_data_2017} However, due to the extremely weak Raman spectral signals, even minor variations can lead to errors in predicting the chemical structure of substances. While this approach may improve model performance to a small extent in specific scenarios, it does not fundamentally resolve the shortage of labeled data. Therefore, effectively utilizing unlabeled data and accurately extracting features to enhance model performance with limited labeled data is of utmost importance. 

In this paper, we propose a spectral self-supervised learning method based on Masked Autoencoders (SMAE), which utilized random masking to disrupt the spectrum as a pretext task for self-supervised training. By reconstructing the original spectrum, we extract effective spectral features, achieving optimal spectral classification performance using only a limited amount of labeled data. SMAE consists of two training phases: a spectral masking self-supervised pretraining phase and a downstream spectral classification fine-tuning task. First, a masked autoencoder is trained to recover the disrupted spectral data, constructing a latent space of internal data representations, with the autoencoder demonstrating effective denoising capabilities. Once the training of the autoencoder is complete, the classification task becomes straightforward, requiring only fine-tuning of network weights with a limited amount of labeled data to achieve excellent spectral recognition performance. Our goal is to establish a flexible model that can extract high-quality chemically relevant information from spectra while accommodating variations in noise and wavelength baseline, enabling learning from different data sources. By revealing the correlations between Raman spectra and omics data, unsupervised learning methods has the potential to transcend conventional understanding of known categorical entities, providing opportunities for hypotheses and generation in fields such as cancer research and personalized medicine\cite{zhang_genotype_2024}. Furthermore, SMAE is not limited to Raman spectroscopy, it can also be applied to other one-dimensional spectral data, such as near-infrared spectroscopy and nuclear magnetic resonance spectroscopy, demonstrating strong scalability.

\section{Methods}
\label{sec:headings}
\begin{figure}
  \centering
  \includegraphics[width=1\linewidth]{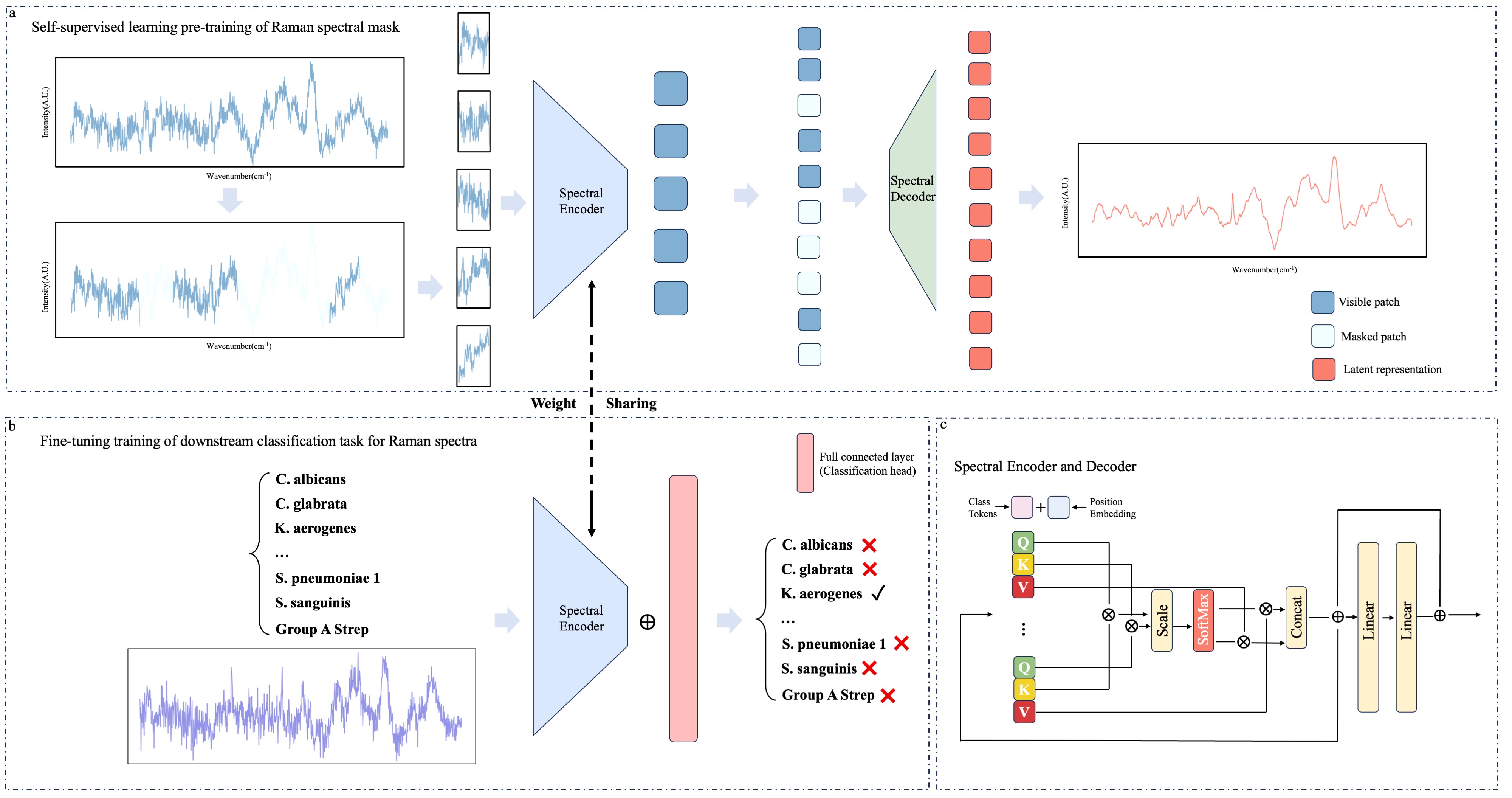}
  \caption{Workflow of the SMAE. (a): Flowchart of the Raman spectrum masked self-supervised learning pre-training phase; (b): Training of the pathogen bacterial classifier using a limited amount of labeled data with shared autoencoder weights; (c): Basic structural block of the multi-head self-attention mechanism in the encoder and decoder.}
  \label{figure_21}
\end{figure}

Our Spectral Masked Autoencoder (SMAE) is a simple autoencoder method aimed at reconstructing partially disrupted spectral signals back into their original complete form, thereby learning latent spectral features. The network must possess an understanding of the spectrum to effectively recover and generate spectral signals. The workflow of SMAE consists of two parts: the self-supervised pretraining phase for spectral masking and the fine-tuning phase for the downstream spectral classification task, as illustrated in Figure \ref{figure_21}. Initially, we pretrain on a large amount of unlabeled data to extract the representation space of the spectra, resulting in an effective encoder. Subsequently, in the downstream classification task, we load the pre-trained encoder weights and only require a small amount of labeled data for model fine-tuning, which significantly improves classification accuracy. This approach not only saves costs associated with labeling spectral data but also greatly reduces computational load. In the following subsection, we will detail each phase.

\subsection{Data sets}
To validate the effectiveness of the proposed SMAE, we selected a large publicly available dataset of pathogenic bacteria (Bacteria-ID)\cite{ho_rapid_2019} to test the feature extraction capability under self-supervised learning and the performance in downstream pathogenic bacterial classification task. The autoencoder demonstrates some denoising capabilities during spectral recovery, and we also performed a denoising effectiveness analysis on the publicly available noisy spectral dataset, the breast cancer cell dataset (MDA-MB-231)\cite{horgan_high-throughput_2021}. The spectral detailed information for both public datasets are shown in the Table \ref{table_21}.

\begin{figure}
  \centering
  \includegraphics[width=1\linewidth]{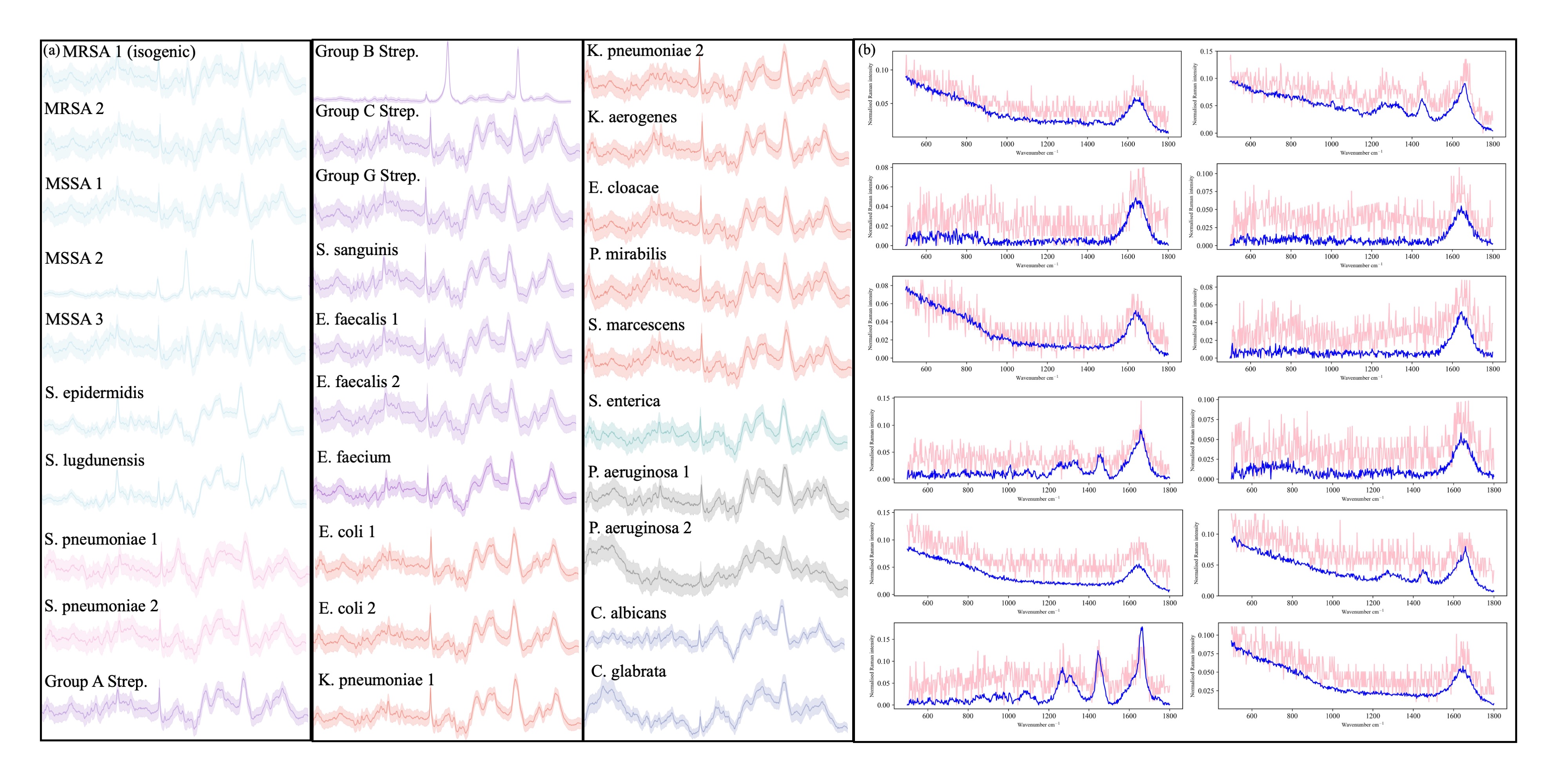}
  \caption{(a) Comparison of spectral differences among 30 species of pathogenic bacteria; (b) Comparison of high and low SNR of spectra in the MDA-MB-231 dataset.}
  \label{figure_22}
\end{figure}

\paragraph{Bacteria-ID:} We utilized the open-source dataset Bacteria-ID for the self-supervised pretraining and fine-tuning validation of the network. This dataset contains Raman spectral data for 30 common pathogenic bacteria, divided into three subsets: reference, finetune, and test, with 2000, 100, and 100 Raman spectral samples per species, respectively. While many studies focus on supervised training, our approach differs by simulating a scenario with no data labels in the reference subset, feeding only spectral data into the network while discarding spectral labels. After training to obtain the representation space, we fine-tuned the classification network using the finetune subset, followed by testing in the test subset to validate the effectiveness of the network’s self-supervised learning. The spectral differences among species are illustrated in Figure \ref{figure_22}(a).

\paragraph{MDA-MB-231:} The MDA-MB-231 dataset contains 159,618 spectral pairs in the training set and 12,694 spectral pairs in the test set, totaling 172,312 spectral pairs. Each pair includes low SNR spectra (with an integration time of 0.1 seconds) and corresponding high SNR spectra (with an integration time of 1 seconds). Data augmentation techniques, such as spectral shifting, flipping, and spectral mixing, are applied to this dataset. We used the low SNR spectra from the training set for pretraining without any corresponding high SNR spectra. The denoising performance of the low SNR spectra was then evaluated in the test set, with the comparison of high and low SNR spectral pairs shown in Figure \ref{figure_22}(b).

\begin{table}
    \centering
  \caption{Brief introduction to the Bacteria-ID and MDA-MB-231 Dataset}
  \label{tbl:example}
  \begin{tabular}{cccc}
    \toprule
    Bacteria-ID\cite{ho_rapid_2019} & Spectra & Classes & Measurement time (s)  \\
    \hline
    Reference  & 60000 & 30 & 1  \\
    Finetune & 3000 & 30 & 1 \\
    Test  & 3000 & 30 & 2 \\
    \bottomrule
     MDA-MB-231\cite{horgan_high-throughput_2021} & & &  \\
    \hline
     Train (pair)  & 159618 & \textbackslash & 0.1/1  \\
     Test (pair) & 12694 & \textbackslash & 0.1/1 \\
    \bottomrule
  \end{tabular}
  \label{table_21}
\end{table}

\subsection{Spectral Self-supervised learning}
The SMAE is a simple and effective paradigm for spectral self-supervised learning. It involves masking a wide range of spectral patches and peaks randomly and reconstructing the original spectrum based on the remaining visible spectral patches. Although extensive masking may result in the loss of important spectral features, the network can still extract complete spectral information when sufficient data is available. The advantage of wide-ranging masking lies in the fact that the missing spectral intensities cannot be recovered through linear fitting with adjacent values, which compels the network to learn more effective features for reconstructing the original spectrum. In the design of the model network for spectral self-supervised pretraining, which falls under the category of generative self-supervised learning, SMAE consists of a simple encoder and decoder. The core design includes creating a pretext task for self-supervised learning by randomly masking the peak intensities of certain wavelengths (e.g., 50\%). The masked spectrum is then fed into the SMAE, with the training objective being to recover it as accurately as possible to the original spectral data.

\subsubsection{Spectral Encoder}
Spectral encoder consists of several Transformer encoder blocks\cite{vaswani_attention_2023}, with the internal structure illustrated in Figure \ref{figure_21}(c). The embedding block of the encoder increases the position encoding through linear mapping, preserving the positional information of the spectral patches while adding a class token in preparation for downstream classification tasks. The encoder processes only the visible spectral patches (e.g., 50\%), while the masked patches are removed and not processed. This reduces the computational load and increases training speed. The complete spectral patches (both visible and masked) will be fed into the decoder, which is described in the next subsection.

\subsubsection{Spectral Decoder}
Spectral decoder is connected to the encoder and is responsible for recovering the original spectral data. It takes in the complete spectral patches (both visible and masked) and reassembles them according to their original positional arrangement, as shown in Figure \ref{figure_21}(a). The decoder is composed of other Transformer encoder structures, utilizing fewer Transformer encoder blocks compared to the encoder, making it more lightweight. It is used solely during pretraining to recover the spectrum and is discarded during downstream classification tasks. The decoder is designed independently from the encoder, providing greater flexibility.

\subsubsection{Spectral Reconstruction Objective}
SMAE reconstructs each masked spectral patches by predicting the intensity of the spectral bands. The output of the decoder provides the intensity for each patch, with the final layer being a linear mapping layer that outputs a number of channels equal to the number of the spectral patches, subsequently reshaping it to the original spectral length. The reconstruction loss function is calculated as the mean squared error (MSE) between the reconstructed and original spectral intensity spaces, with the loss computed only for the masked spectral patch intensities. In the initial experiments validating the spectral reconstruction performance, training was first conducted on the Bacteria-ID dataset, which generally has low SNR. However, the newly reconstructed spectra not only effectively recover the original spectral features but also significantly reduce spectral noise, demonstrating a certain degree of denoising capability. This indicates that the pretrained encoder possesses an excellent understanding of spectral features. The results and discussion section will present the spectral reconstruction performance and analyze the denoising capability.

\subsection{Fine-Tuning Spectral Classification Task}
In scenarios where labeling measured spectral data is difficult and time-consuming, significant improvements in classification accuracy can be achieved using a self-supervised learning approach with only a minimal amount of labeled data. After completing the masked self-supervised pretraining, the encoder weights are shared, and a fully connected layer for classification is added after the encoder, with the number of output neurons corresponding to the final number of classes, thus completing the spectral classification task. The network design is extremely simple and yield favorable results, as illustrated in the structure shown in Figure \ref{figure_21}(b).

\section{Results and discussion}
\subsection{Spectral Reconstruction and Denoising}
\begin{figure}
  \centering
  \includegraphics[width=0.65\linewidth]{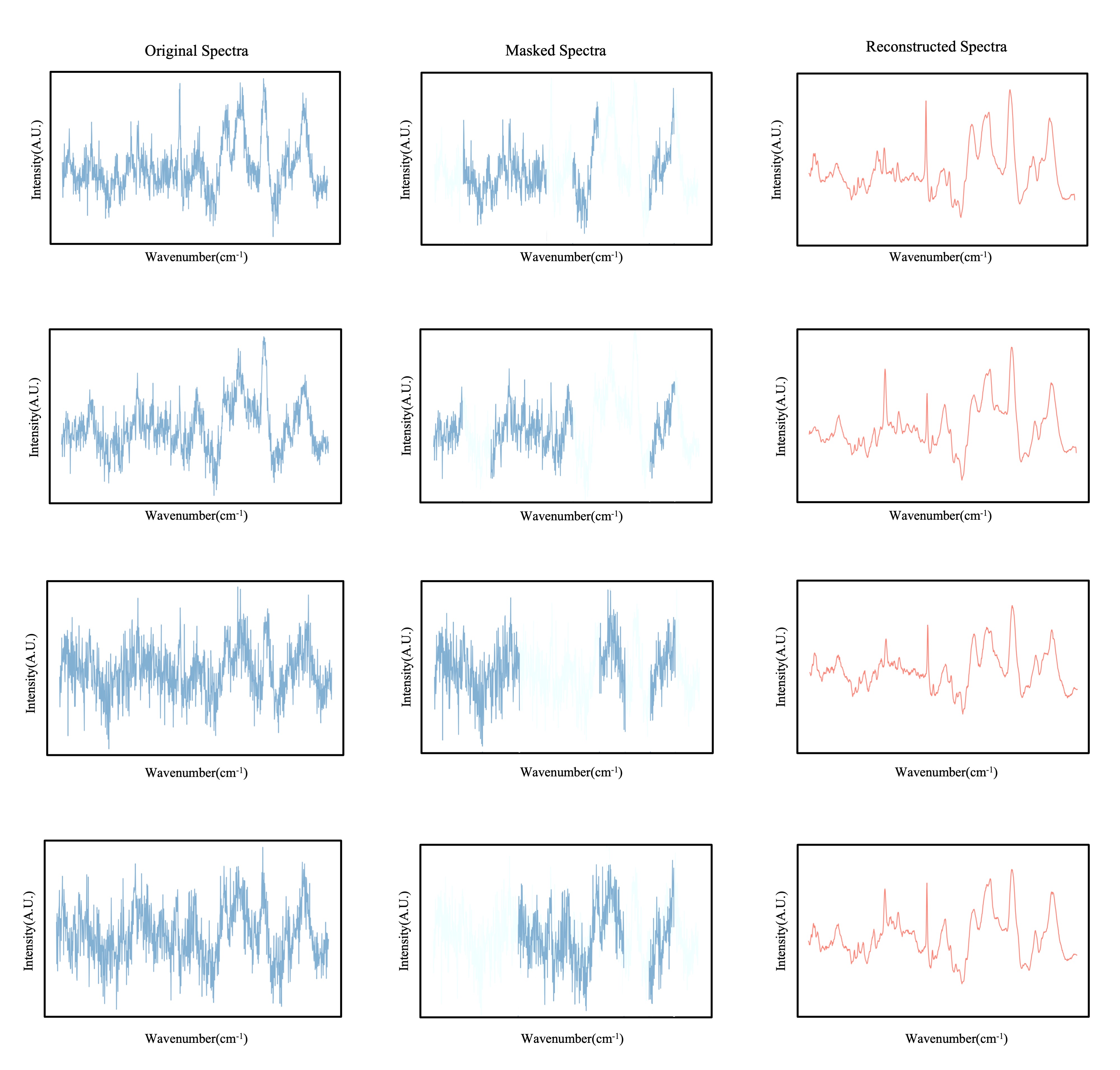}
  \caption{The Bacteria-ID test spectral data, the left, middle, and right columns represent the original spectral plots, masked spectral plots, and reconstructed spectral plots, respectively.}
  \label{figure_31}
\end{figure}

The effectiveness of spectral reconstruction reflects the network’s understanding of spectra and the encoder’s ability to extract spectral features, which impacts the classification performance after fine-tuning the weights in downstream tasks. We pretrained the model on the reference subset of the Bacteria-ID dataset, and after pretraining, we input spectra from the test subset, again applying random spectral masking to reconstruct the original unmasked spectral data. The original spectra, masked spectra and reconstructed spectra are shown in Figure \ref{figure_31}. It is noted that the reconstructed spectra, to some extent, eliminate spectral noise while retaining important spectral peaks, with the recovered peak features becoming more highlighted.

\begin{figure}
  \centering
  \includegraphics[width=0.75\linewidth]{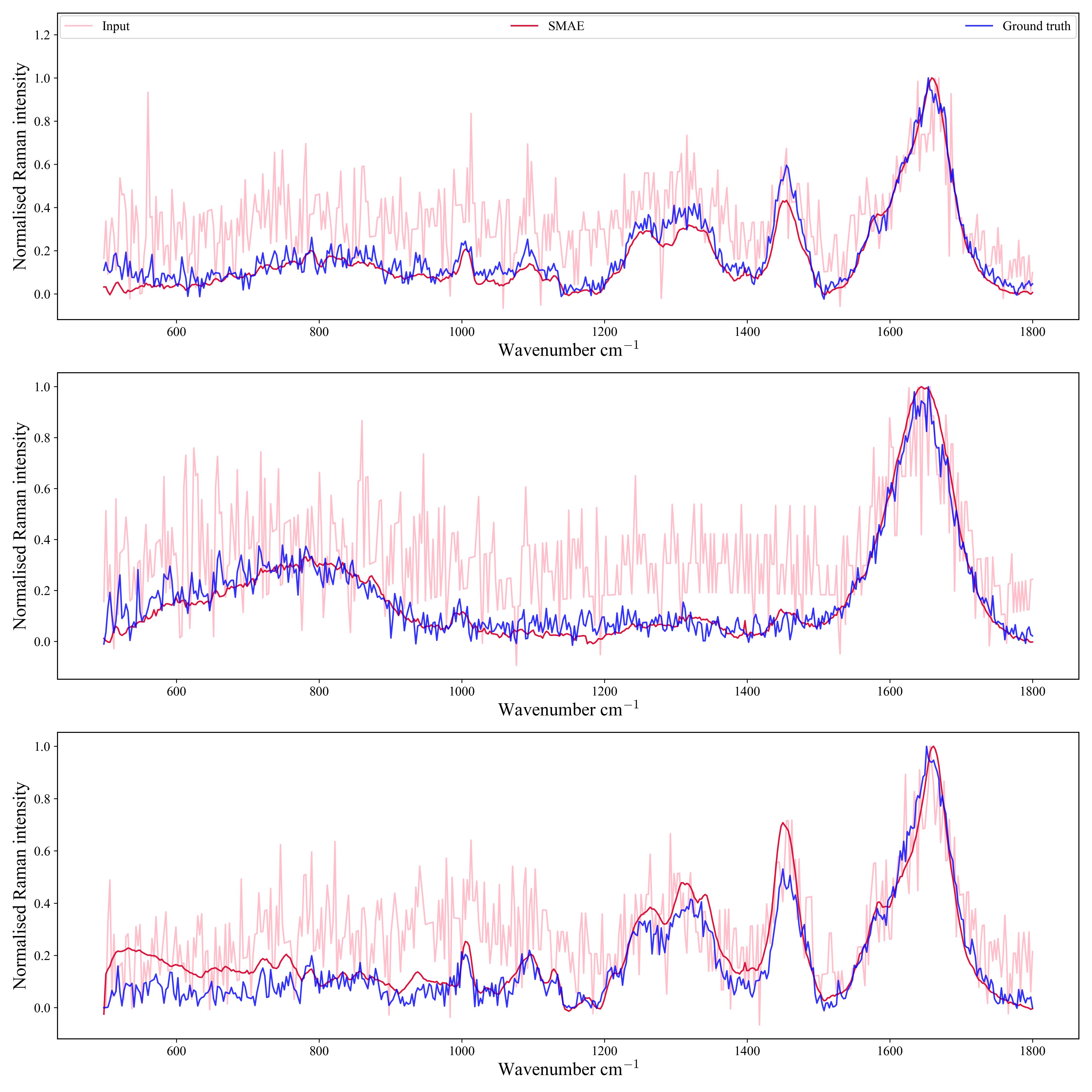}
  \caption{Comparison of denoising performance of SMAE for three randomly selected test spectra.}
  \label{figure_32}
\end{figure}

To validate the denoising capability of SMAE, we first conducted pretraining on the MDA-MB-231 breast cancer cell dataset, masking 50\% of the training data without using any high SNR spectra as training labels. After pretraining, we reconstructed the low SNR spectra in the test set, achieving more than a twofold increase in SNR compared to the original spectra. Table \ref{table_31} presents a comparison of the original spectra and the evaluation metrics after denoising, while Figure \ref{figure_32} compares three randomly selected low SNR spectra, high SNR spectra, and the reconstructed spectra from SMAE. The reconstructed spectra effectively retained the characteristic peaks and smoothed out spectral noise, demonstrating a certain degree of denoising capability.

\begin{table}
    \centering
  \caption{Comparison of denoising evaluation metrics in the MDA-MB-231 Dataset.}
  \label{tbl:example}
  \begin{tabular}{cccc}
    \toprule
     MDA-MB-231 & Original spectra & Without SMAE pretraining & SMAE pretraining \\
    \hline
    SNR  & 5.0883 & 1.8218 & 10.4039  \\
    MSE & 0.0425 & 0.1093 & 0.0125 \\
     \bottomrule
  \end{tabular}
  \label{table_31}
\end{table}

\subsection{Feature Extraction Capability}
\begin{figure}
  \centering
  \includegraphics[width=1\linewidth]{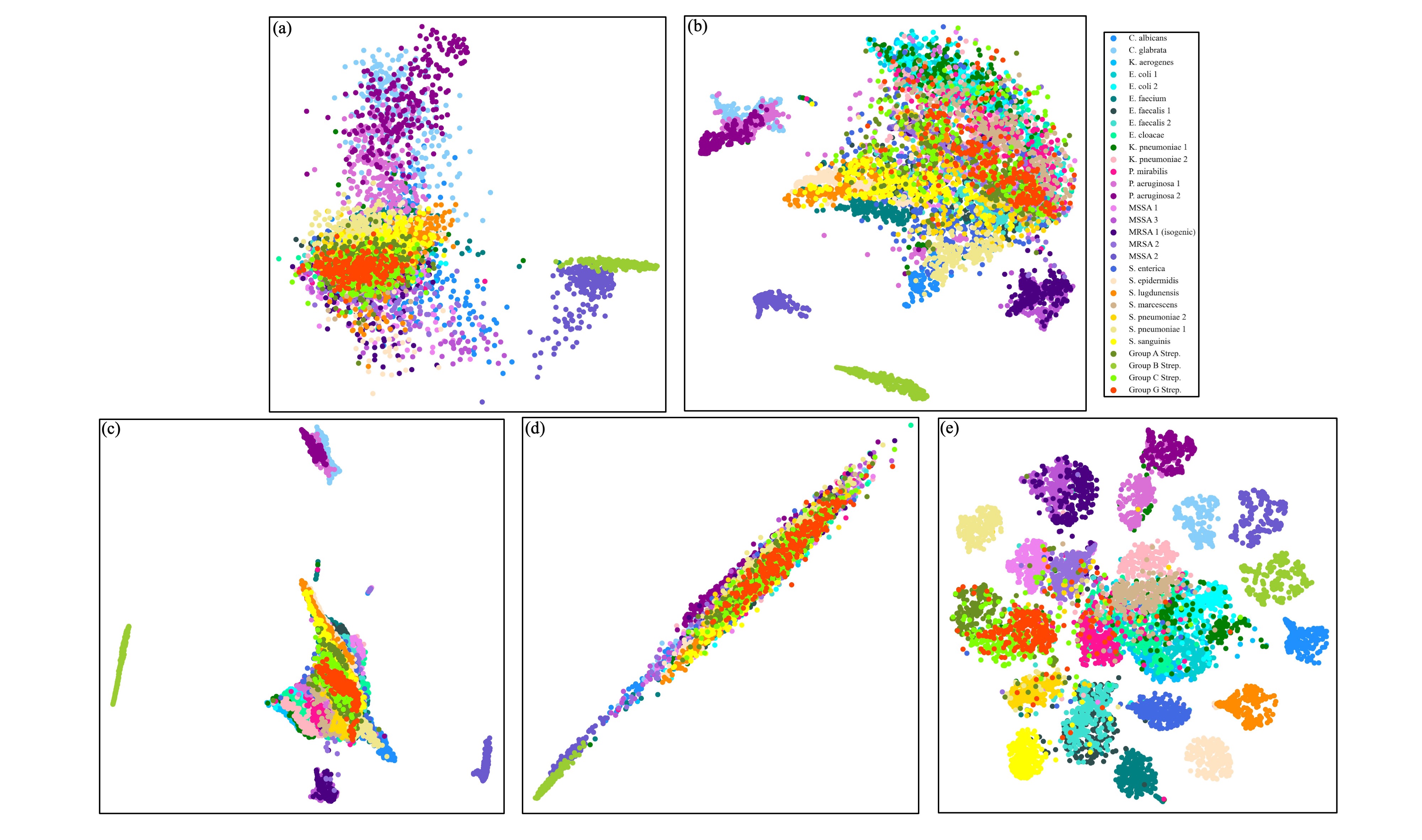}
  \caption{Clustering comparison diagram for unsupervised learning, where (a), (b), (c), (d) and (e) represent PCA, t-SNE, UMAP, SOM, and SMAE, respectively.}
  \label{figure_33}
\end{figure}

To validate the feature extraction performance of SMAE, we compared our method with four classical unsupervised feature extraction techniques: PCA, t-SNE, Uniform Manifold Approximation and Projection (UMAP), and Self-organizing map (SOM). All methods were applied to reduce the dimensionality of features in the reference subset of Bacteria-ID dataset, followed by K-means clustering to achieve classification for 30 species. It is noteworthy that no labeled data was utilized. The results presented in the Table \ref{table_32} demonstrate that SMAE shows a significant improvement in classification accuracy compared to classical unsupervised methods. Figure \ref{figure_33} illustrates the K-means clustering results after feature extraction for the five methods, indicating that SMAE can extract more highlighted features in unsupervised learning. Due to the large number of categories and the volume of spectral data, classical methods struggled to cluster effectively. Although some sample overlap exists in SMAE, the extracted features yielded the best clustering performance.

\begin{table}
    \centering
  \caption{Comparison of clustering results between SMAE and classical unsupervised methods.}
  \label{tbl:example}
  \begin{tabular}{cc}
    \toprule
     Methods & Clustering Accuracy \\
    \hline
    K-means  & 37.86\% \\
    PCA+K-means  & 19.97\% \\
    t-SNE+K-means  & 33.73\% \\
    UMAP+K-means  & 37.05\% \\
    SOM+K-means & 16.08\% \\
    \textbf{SMAE+K-means} & \textbf{80.56\%} \\
     \bottomrule
  \end{tabular}
  \label{table_32}
\end{table}

\begin{figure}
  \centering
  \includegraphics[width=0.75\linewidth]{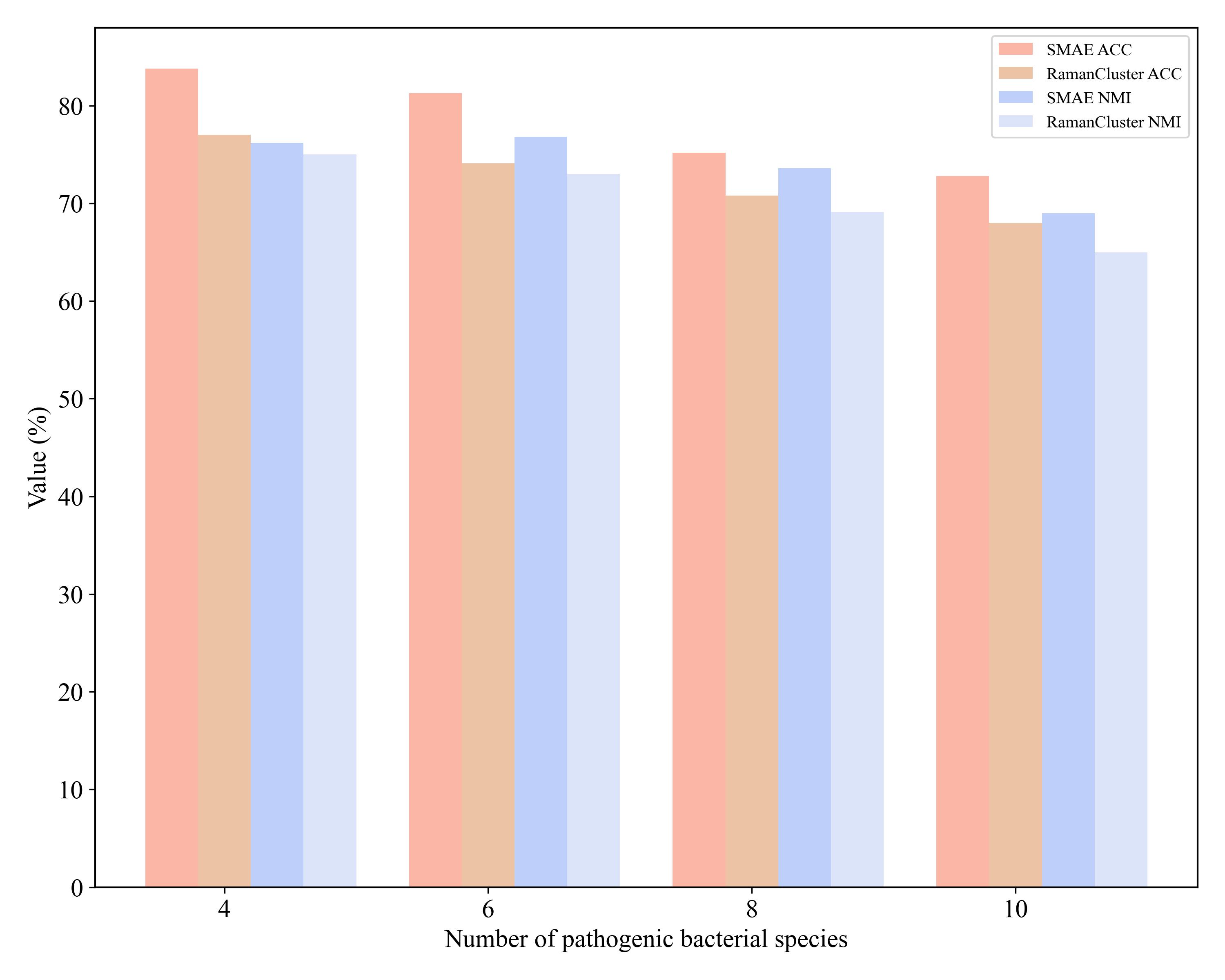}
  \caption{Comparison of classification performance between SMAE and RamanCluster for pathogen bacteria species counts of 4, 6, 8, and 10.}
  \label{figure_34}
\end{figure}

In the comparison of deep clustering methods, we followed the dataset partitioning strategy outlined in RamanCluster\cite{sun_ramancluster_2024} to test the classification accuracy of unsupervised clustering in Bacteria-4 and Bacteria-6. Additionally, we compared the clustering performance of SMAE with four other state-of-the-art deep clustering approaches such as SimCLR\cite{chen_simple_2020}, Contrastive Clustering (CC)\cite{li_contrastive_2020}, TS-TCC\cite{eldele_time-series_2021}, and Supporting Clustering with Contrastive learning (SCCL)\cite{zhang_supporting_2021}. SMAE demonstrated an improvement of over 6\% in clustering accuracy compared to RamanCluster, which was the method with the highest accuracy. Moreover, SMAE showed improvements in both Normalized Mutual Information (NMI) and Adjusted Mutual Information (AMI). The comparison results of the relevant evaluation metrics are presented in the Table \ref{table_33}. We further increased the number of sample species to Bacteria-8 and Bacteria-10, with the test results illustrated in the Figure \ref{figure_34}. Compared to RamanCluster, SMAE demonstrated outstanding feature extraction performance even with a larger number of species.

\begin{table}
    \centering
  \caption{Comparison of SMAE with five other deep clustering methods.}
  \label{tbl:example}
  \begin{tabular}{ccccccc}
   	\toprule
	\multirow{2}{*}{Mehthods} & \multicolumn{3}{c}{Bacteria-4} &\multicolumn{3}{c}{Bacteria-6} \\
	\cmidrule{2-7}
            & ACC & NMI & AMI &ACC & NMI & AMI \\
        \midrule
        
        SimCLR\cite{chen_simple_2020} & 65.3\% & 54.4\% & 54.4\% & 53.5\% & 47.6\% & 47.4\% \\
        SCCL\cite{zhang_supporting_2021} & 71.0\% & 61.7\% & 61.6\% & 52.0\% & 46.4\% & 46.2\% \\
        CC\cite{li_contrastive_2020} & 71.6\% & 61.0\% & 60.8\% & 54.7\% & 53.5\% & 53.3\% \\
        TS-TCC\cite{eldele_time-series_2021} & 75.0\% & 73.5\% & 73.4\% & 70.5\% & 65.7\% & 64.8\% \\
        RamanCluster\cite{sun_ramancluster_2024} & 77.0\% & 75.0\% & 74.6\% & 74.1\% & 73.0\% & 72.6\% \\
        \textbf{SMAE} & \textbf{83.8\%} & \textbf{76.2\%} & \textbf{76.1\%} & \textbf{81.3\%} & \textbf{76.8\%} & \textbf{76.7\%} \\
        \bottomrule
  \end{tabular}
  \label{table_33}
\end{table}

\subsection{Performance of Downstream Classification Tasks}
\begin{figure}
  \centering
  \includegraphics[width=1\linewidth]{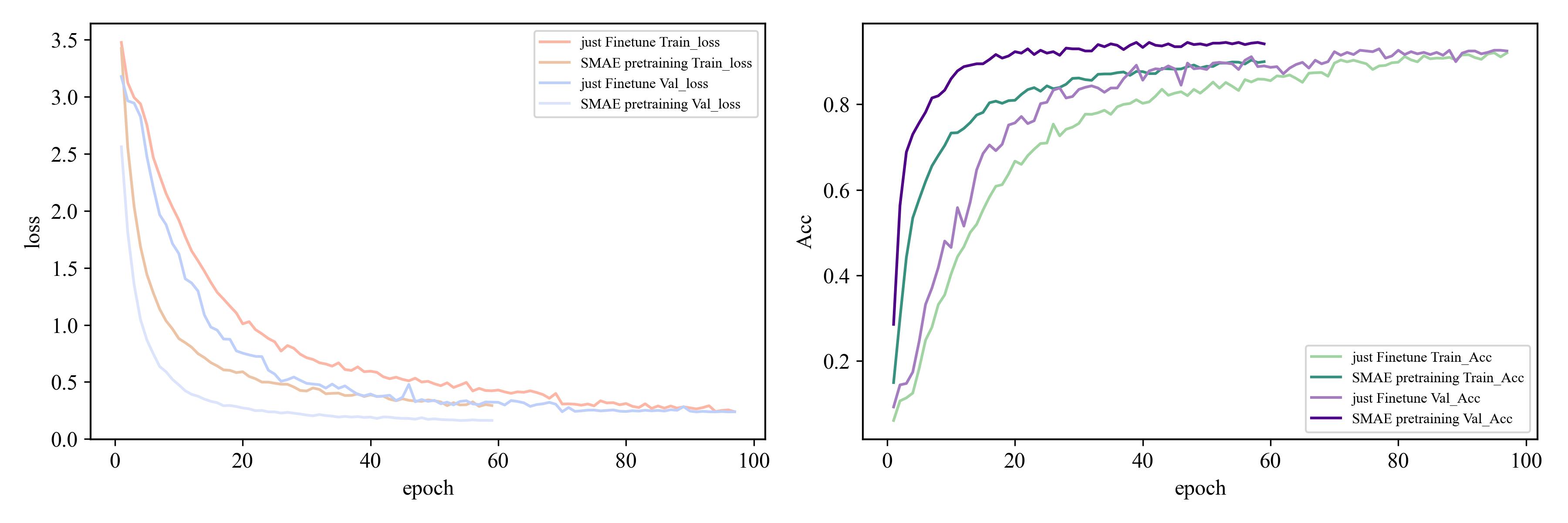}
  \caption{Loss and accuracy variation curves for SMAE with and without pretraining.}
  \label{figure_35}
\end{figure}

\begin{figure}
  \centering
  \includegraphics[width=1\linewidth]{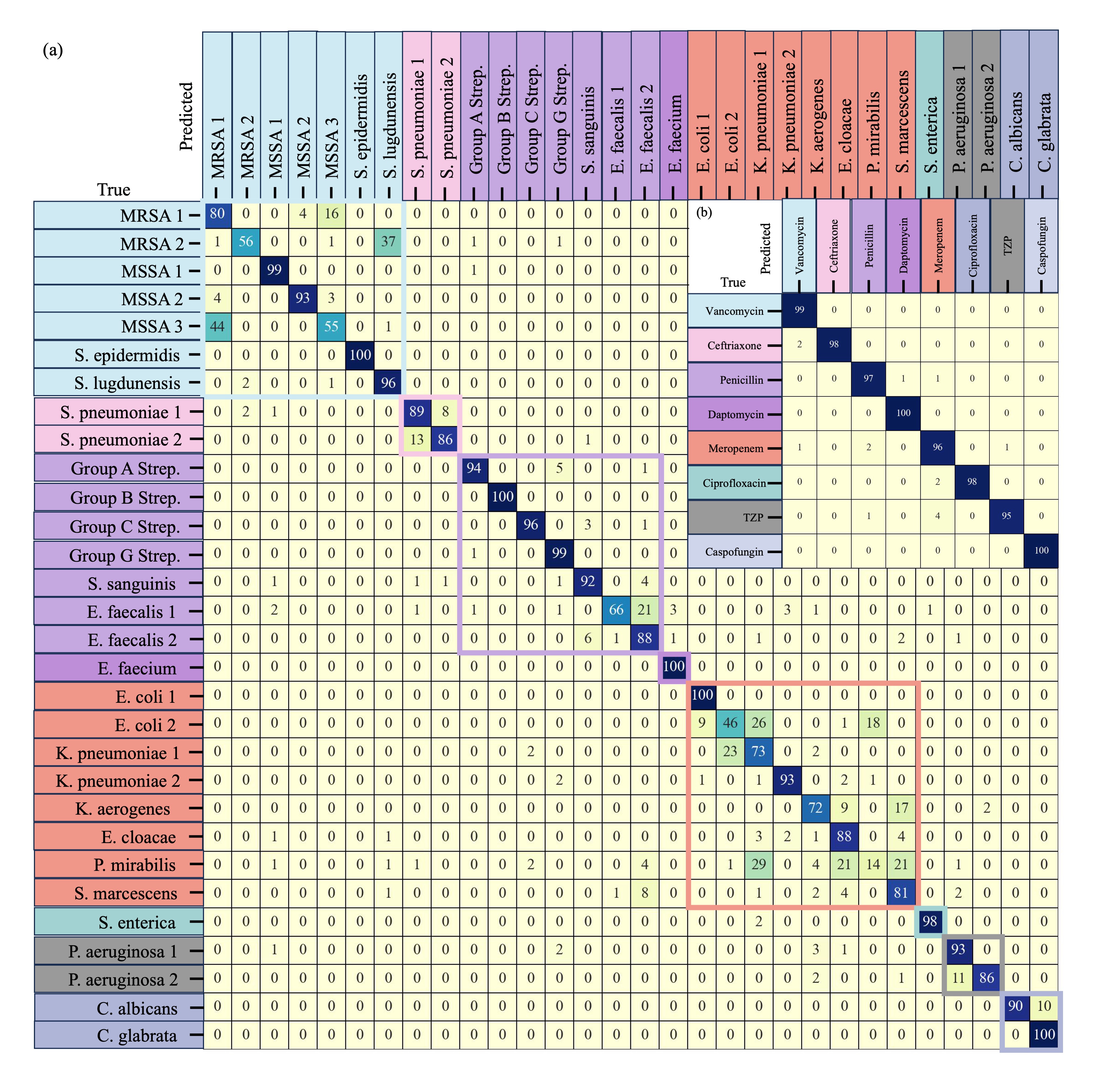}
  \caption{Confusion matrices of (a) 30-isolates and (b) 8-empiric-treatments.}
  \label{figure_36}
\end{figure}

SMAE was pretrained on the reference subset of Bacteria-ID dataset, utilizing the finetune subset to adjust the pretrained encoder weights, which were then tested on the test subset. SMAE achieved better classification performance in the reference subset without using any data labels, compared to training with only the finetune subset (77.80\%). After pretraining and subsequent fine-tuning, there was an increase of over 6\% in accuracy (83.90\%). The loss and accuracy curves for training and validation sets when trained only with the fine-tuning subset versus after loading pretrained weights are shown in the Figure \ref{figure_35}, indicating rapid convergence for both loss and accuracy. This approach addressed the challenge of effective feature extraction in the absence of data labels, achieving competitive classification results for pathogens compared to the supervised ResNet (83.40\%), and the comparison results are presented in Table \ref{table_34}. In the Figure \ref{figure_36}, confusion matrices (a and b) illustrate the classification details for 30 types of isolates and 8 types of empirical treatment tasks.

\begin{table}
    \centering
  \caption{Comparison of SMAE with supervised learning ResNet methods.}
  \label{tbl:example}
  \begin{tabular}{cccc}
    \toprule
     Methods & \makecell[c]{Supervised learning\\ Accuracy} & \makecell[c]{w/o pretraining\\ Accuracy} & \makecell[c]{w/ pretraining\\ Accuracy} \\
    \hline
    ResNet  & 83.40\% & 76.50\% & \textbackslash \\
    SMAE & 84.80\% & 77.80\% & 83.90\% \\
     \bottomrule
  \end{tabular}
  \label{table_34}
\end{table}

\begin{figure}
  \centering
  \includegraphics[width=1\linewidth]{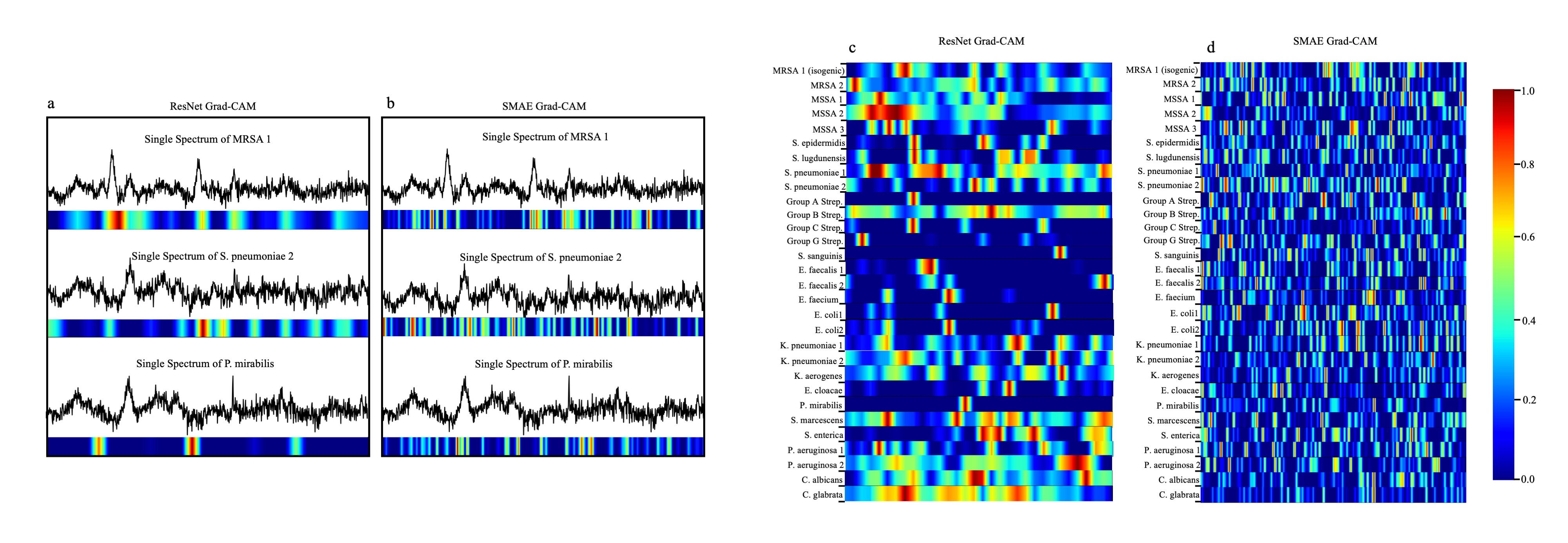}
  \caption{Grad-CAM applied to pathogenic bacterial classification using ResNet and SMAE: (a) and (b) Random selection of 3 species (MRSA1, S. pneumoniae2, P. mirabilis); (c) and (d) Stacked test set (30-isolate).}
  \label{figure_37}
\end{figure}

To further visualize the feature extraction process of self-supervised learning, we introduced gradient-weighted class activation mapping (Grad-CAM)\cite{selvaraju_grad-cam_2017}, which helps explain the model’s high-performance transfer capability and enhances the interpretability of the neural network. Grad-CAM is applied after the encoder structure, allowing us to identify the key features that are crucial for the classification results derived from the network. In Figure \ref{figure_37}(a) and (b) present the spectral classification weight visualization maps for ResNet and SMAE, respectively, across three pathogenic bacterial species (MRSA1, S. pneumoniae2, P. mirabilis). This highlights the wavelength information that plays a decisive role in classification, with relevant features depicted in warm colors and irrelevant features in cool colors. To comprehensively illustrate the differences in features extracted by ResNet and SMAE across 30 pathogenic bacterial species, Figure \ref{figure_37}(c) and (d) show the complete classification weight visualization maps for all thirty species. ResNet primarily extracts features through convolutional operations, yielding prominent features in large local spectral bands. In contrast, SMAE’s main structure is a multi-head self-attention module that focuses more on the interrelationships among spectral patches, thus enhancing the extraction of subtle global features. The self-supervised learning approach, which does not utilize labels as learning objectives, enables the discovery of finer differences among various bacterial populations. This methodology can serve as a reference for further analysis of the pathogenic Raman spectra of biomolecules in the future.

\subsection{Discussion of Main Properties}
The primary innovation of SMAE is the introduction of a spectral masking strategy to learn latent features. Consequently, the masking ratio of the spectrum is crucial in fine-tuning classification tasks. A high masking ratio leads to the loss of significant spectral peak information, while a low masking ratio may cause the network to become complacent, making the reconstruction of spectra overly simplistic and hindering the learning of useful information. Additionally, the size of the spectral patches, as well as the architectures of encoder and decoder, will also impact the effectiveness of feature learning. The following section will discuss the main properties of the SMAE.

\begin{figure}
  \centering
  \includegraphics[width=1\linewidth]{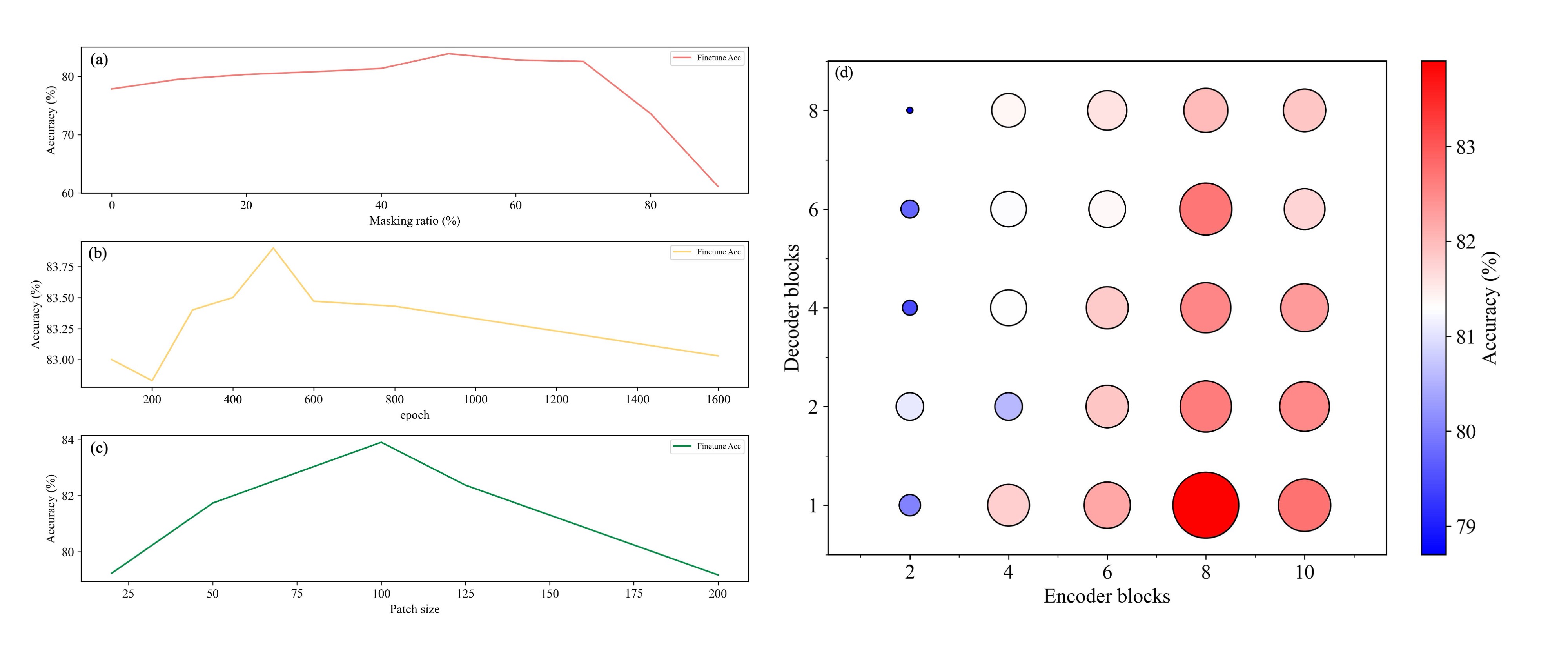}
  \caption{The impact of main properties setting of SMAE on classification accuracy: Subplots (a), (b), (c), and (d) correspond to the masking ratio, training epochs, patch size, and the structure of the encoder and decoder, respectively.}
  \label{figure_38}
\end{figure}

\paragraph{Masking Ratio:} When the masking ratio is set to 0\%, meaning that no processing is applied to the input spectra, SMAE merely re-encodes the spectra. In the absence of pretext task, the model fails to learn any effective features, resulting in a fine-tuning accuracy of only 78.40\%. At masking ratio of 50\%, SMAE exhibits the best feature learning capability, achieving a fine-tuning accuracy of 83.90\%. However, when the masking ratio exceeds 80\%, a significant loss of spectral information occurs, causing the accuracy to drop to around 77\%, which is comparable to the results obtained using only the finetune subset for training. After determining the optimal pre-training masking ratio of 50\%, we evaluate the training epochs. When trained for 500 epochs, the fine-tuning classification accuracy reaches its peak, and further training does not yield a significant improvement in model performance. The result of the classification accuracy variations with different masking ratio and training epochs are illustrated in Figure \ref{figure_38}(a) and (b).

\paragraph{Patch size:} In the multi-head self-attention module of the spectra, the size of the spectral patches is critically important. Smaller spectral patches may lead to overfitting in the model, while larger patches can result in the loss of local detail information, consequently decreasing accuracy. We tested patch sizes of 20, 50, 100, 125, and 200, respectively. The results indicate that when the spectral dimension is 1000, a patch size of 100 yields the best fine-tuning accuracy, effectively preserving more spectral detail features. The influence of patch size on fine-tuning accuracy is illustrated in the Figure \ref{figure_38}(c).

\paragraph{Depth of Spectral Encoder and Decoder:} SMAE consists of several multi-head self-attention modules. The decoder’s exclusive function is to reconstruct the spectrum, and a deeper decoder might negatively impact the feature extraction capability of the encoder. Therefore, we tested different depths of both the encoder (with 2, 4, 6, 8, and 10 blocks) and the decoder (with 1, 2, 4, 6, and 8 blocks). The results show that the optimal fine-tuning classification accuracy is achieved when the encoder consists of 8 multi-head self-attention modules and the decoder consists of 1. When the decoder is deeper than the encoder, the fine-tuning classification accuracy tends to decrease, as the decoder diminishes the feature extraction capability of the encoder. The influence of the encoder and decoder structures on fine-tuning accuracy is shown in the Figure \ref{figure_38}(d).

\section{Conclusions}
To address the challenge of feature extraction from unlabeled spectra, we propose a simple yet effective self-supervised learning paradigm for Raman spectroscopy, termed SMAE. SMAE proposed a learning strategy for random masked spectra for the first time, which stimulated the potential of self-supervised learning in spectral feature extraction. SMAE consists of spectral self-supervised pre-training and fine-tuning for downstream classification tasks. We validate the denoising and classification performance of SMAE using the MDA-MB-231 breast cancer cell dataset and the Bacteria-ID pathogen bacteria dataset. After pre-training, the reconstructed spectra effectively smooth out numerous spikes, making spectral peaks more distinct and improving the SNR by more than twofold. In spectral classification without any labels, leveraging the pre-trained network weights and using t-SNE and K-means for dimensionality reduction and clustering, SMAE achieves an accuracy of over 80\% in identifying 30 pathogen bacterial species, significantly outperforming classical clustering methods and other state-of-the-art deep clustering methods. After fine-tuning with a limited amount of labeled data, SMAE further improves the classification accuracy of pathogen bacteria to 83.90\% on the test set, demonstrating competitive performance compared to the supervised ResNet (83.40\%). This work visualizes the feature extraction process in self-supervised learning for spectra, enhancing the interpretability of the network. Furthermore, we discuss the impact of main properties settings in SMAE on classification accuracy. Self-supervised learning offers certain advantages over supervised learning, as it does not rely on data labels and can learn feature beyond those typically captured by supervised methods, enabling the exploration of more detailed chemical information from spectra. Self-supervised learning is poised to become a powerful tool for spectral feature learning.

\section*{Acknowledgement}
This work is supported by the National Key R\&D Plan under Grant No. 2021YFF0601200 and 2021YFF0601204.

\bibliographystyle{unsrt}  
\bibliography{SMAE}

\begin{thebibliography}{10}

\bibitem{cheng_vibrational_2015}
Ji-Xin Cheng and X.~Sunney Xie.
\newblock Vibrational spectroscopic imaging of living systems: {An} emerging platform for biology and medicine.
\newblock {\em Science}, 350(6264):aaa8870, November 2015.

\bibitem{wang_multi-point_2023}
Yu~Wang, Hao Peng, Kunxiang Liu, Lindong Shang, Lei Xu, Zhenming Lu, and Bei Li.
\newblock Multi-point scanning confocal {Raman} spectroscopy for accurate identification of microorganisms at the single-cell level.
\newblock {\em Talanta}, 254:124112, March 2023.

\bibitem{kamp_raman_2024}
Marlous Kamp, Jakub Surmacki, Marc Segarra~Mondejar, Tim Young, Karolina Chrabaszcz, Fadwa Joud, Vincent Zecchini, Alyson Speed, Christian Frezza, and Sarah~E. Bohndiek.
\newblock Raman micro-spectroscopy reveals the spatial distribution of fumarate in cells and tissues.
\newblock {\em Nature Communications}, 15(1):5386, June 2024.

\bibitem{guselnikova_pretreatment-free_2024}
Olga Guselnikova, Andrii Trelin, Yunqing Kang, Pavel Postnikov, Makoto Kobashi, Asuka Suzuki, Lok~Kumar Shrestha, Joel Henzie, and Yusuke Yamauchi.
\newblock Pretreatment-free {SERS} sensing of microplastics using a self-attention-based neural network on hierarchically porous {Ag} foams.
\newblock {\em Nature Communications}, 15(1):4351, May 2024.

\bibitem{zhang_alkyne-tagged_2024}
Sihan Zhang, Yuxiao Mei, Jiaqi Liu, Zhichao Liu, and Yang Tian.
\newblock Alkyne-tagged {SERS} nanoprobe for understanding {Cu}+ and {Cu2}+ conversion in cuproptosis processes.
\newblock {\em Nature Communications}, 15(1):3246, April 2024.

\bibitem{shin_single_2023}
Hyunku Shin, Byeong~Hyeon Choi, On~Shim, Jihee Kim, Yong Park, Suk~Ki Cho, Hyun~Koo Kim, and Yeonho Choi.
\newblock Single test-based diagnosis of multiple cancer types using {Exosome}-{SERS}-{AI} for early stage cancers.
\newblock {\em Nature Communications}, 14(1):1644, March 2023.

\bibitem{huang_rapid_2023}
Liping Huang, Hongwei Sun, Liangbin Sun, Keqing Shi, Yuzhe Chen, Xueqian Ren, Yuancai Ge, Danfeng Jiang, Xiaohu Liu, Wolfgang Knoll, Qingwen Zhang, and Yi~Wang.
\newblock Rapid, label-free histopathological diagnosis of liver cancer based on {Raman} spectroscopy and deep learning.
\newblock {\em Nature Communications}, 14(1):48, January 2023.

\bibitem{bresci_label-free_2024}
Arianna Bresci, Koseki~J. Kobayashi-Kirschvink, Giulio Cerullo, Renzo Vanna, Peter T.~C. So, Dario Polli, and Jeon~Woong Kang.
\newblock Label-free morpho-molecular phenotyping of living cancer cells by combined {Raman} spectroscopy and phase tomography.
\newblock {\em Communications Biology}, 7(1):785, June 2024.

\bibitem{xue_advances_2023}
Xi~Xue, Hanyu Sun, Minjian Yang, Xue Liu, Hai-Yu Hu, Yafeng Deng, and Xiaojian Wang.
\newblock Advances in the {Application} of {Artificial} {Intelligence}-{Based} {Spectral} {Data} {Interpretation}: {A} {Perspective}.
\newblock {\em Analytical Chemistry}, page acs.analchem.3c02540, September 2023.

\bibitem{lu_patch-based_2024}
Xin-Yu Lu, Chen-Yue Wang, Hui Tang, Yi-Fei Qin, Li~Cui, Xiang Wang, Guo-Kun Liu, and Bin Ren.
\newblock Patch-{Based} {Convolutional} {Encoder}: {A} {Deep} {Learning} {Algorithm} for {Spectral} {Classification} {Balancing} the {Local} and {Global} {Information}.
\newblock {\em Analytical Chemistry}, page acs.analchem.3c03889, February 2024.

\bibitem{chen_combined_2024}
Junfan Chen, Jiaqi Hu, Chenlong Xue, Qian Zhang, Jingyan Li, Ziyue Wang, Jinqian Lv, Aoyan Zhang, Hong Dang, Dan Lu, Defeng Zou, Longqing Cong, Yuchao Li, Gina~Jinna Chen, and Perry~Ping Shum.
\newblock Combined {Mutual} {Learning} {Net} for {Raman} {Spectral} {Microbial} {Strain} {Identification}.
\newblock {\em Anal. Chem.}, 2024.

\bibitem{deng_scale-adaptive_2022}
Lin Deng, Yuzhong Zhong, Maoning Wang, Xiujuan Zheng, and Jianwei Zhang.
\newblock Scale-{Adaptive} {Deep} {Model} for {Bacterial} {Raman} {Spectra} {Identification}.
\newblock {\em IEEE Journal of Biomedical and Health Informatics}, 26(1):369--378, January 2022.

\bibitem{he_deep_2021}
Hao He, Sen Yan, Danya Lyu, Mengxi Xu, Ruiqian Ye, Peng Zheng, Xinyu Lu, Lei Wang, and Bin Ren.
\newblock Deep {Learning} for {Biospectroscopy} and {Biospectral} {Imaging}: {State}-of-the-{Art} and {Perspectives}.
\newblock {\em Anal. Chem.}, 2021.

\bibitem{ma_classifying_2021}
Danying Ma, Linwei Shang, Jinlan Tang, Yilin Bao, Juanjuan Fu, and Jianhua Yin.
\newblock Classifying breast cancer tissue by {Raman} spectroscopy with one-dimensional convolutional neural network.
\newblock {\em Spectrochimica Acta Part A: Molecular and Biomolecular Spectroscopy}, 256:119732, July 2021.

\bibitem{liu_dynamic_2019}
Jinchao Liu, Stuart~J. Gibson, James Mills, and Margarita Osadchy.
\newblock Dynamic spectrum matching with one-shot learning.
\newblock {\em Chemometrics and Intelligent Laboratory Systems}, 184:175--181, January 2019.

\bibitem{sun_k-means_2020}
Yi~Sun, Ethan~W. Chen, Jalen Thomas, Yuan Liu, Haohua Tu, and Stephen~A. Boppart.
\newblock K-means clustering of coherent {Raman} spectra from extracellular vesicles visualized by label-free multiphoton imaging.
\newblock {\em Optics Letters}, 45(13):3613, July 2020.

\bibitem{arslan_discrimination_2022}
Afra~Hacer Arslan, Fatma~Uysal Ciloglu, Ummugulsum Yilmaz, Emrah Simsek, and Omer Aydin.
\newblock Discrimination of waterborne pathogens, {Cryptosporidium} parvum oocysts and bacteria using surface-enhanced {Raman} spectroscopy coupled with principal component analysis and hierarchical clustering.
\newblock {\em Spectrochimica Acta Part A: Molecular and Biomolecular Spectroscopy}, 267:120475, February 2022.

\bibitem{uysal_ciloglu_identification_2020}
Fatma Uysal~Ciloglu, Ayse~Mine Saridag, Ibrahim~Halil Kilic, Mahmut Tokmakci, Mehmet Kahraman, and Omer Aydin.
\newblock Identification of methicillin-resistant \textit{{Staphylococcus} aureus} bacteria using surface-enhanced {Raman} spectroscopy and machine learning techniques.
\newblock {\em The Analyst}, 145(23):7559--7570, 2020.

\bibitem{wang_applications_2021}
Liang Wang, Wei Liu, Jia-Wei Tang, Jun-Jiao Wang, Qing-Hua Liu, Peng-Bo Wen, Meng-Meng Wang, Ya-Cheng Pan, Bing Gu, and Xiao Zhang.
\newblock Applications of {Raman} {Spectroscopy} in {Bacterial} {Infections}: {Principles}, {Advantages}, and {Shortcomings}.
\newblock {\em Frontiers in Microbiology}, 12:683580, July 2021.

\bibitem{guo_unsupervised_2022}
Yixin Guo, Weiqi Jin, Weilin Wang, Zongyu Guo, and Yuqing He.
\newblock Unsupervised convolutional variational autoencoder deep embedding clustering for {Raman} spectra.
\newblock {\em Analytical Methods}, 14(39):3898--3910, 2022.

\bibitem{sun_ramancluster_2024}
Zhijian Sun, Zhuo Wang, and Mingqi Jiang.
\newblock {RamanCluster}: {A} deep clustering-based framework for unsupervised {Raman} spectral identification of pathogenic bacteria.
\newblock {\em Talanta}, 275:126076, August 2024.

\bibitem{balestriero_cookbook_2023-1}
Randall Balestriero, Mark Ibrahim, Vlad Sobal, Ari Morcos, Shashank Shekhar, Tom Goldstein, Florian Bordes, Adrien Bardes, Gregoire Mialon, Yuandong Tian, Avi Schwarzschild, Andrew~Gordon Wilson, Jonas Geiping, Quentin Garrido, Pierre Fernandez, Amir Bar, Hamed Pirsiavash, Yann LeCun, and Micah Goldblum.
\newblock A {Cookbook} of {Self}-{Supervised} {Learning}, June 2023.
\newblock arXiv:2304.12210 [cs].

\bibitem{he_masked_2022}
Kaiming He, Xinlei Chen, Saining Xie, Yanghao Li, Piotr Dollar, and Ross Girshick.
\newblock Masked {Autoencoders} {Are} {Scalable} {Vision} {Learners}.
\newblock In {\em 2022 {IEEE}/{CVF} {Conference} on {Computer} {Vision} and {Pattern} {Recognition} ({CVPR})}, pages 15979--15988, New Orleans, LA, USA, June 2022. IEEE.

\bibitem{xie_simmim_2022}
Zhenda Xie, Zheng Zhang, Yue Cao, Yutong Lin, Jianmin Bao, Zhuliang Yao, Qi~Dai, and Han Hu.
\newblock {SimMIM}: {A} {Simple} {Framework} for {Masked} {Image} {Modeling}, April 2022.
\newblock arXiv:2111.09886 [cs].

\bibitem{chen_context_2024}
Xiaokang Chen, Mingyu Ding, Xiaodi Wang, Ying Xin, Shentong Mo, Yunhao Wang, Shumin Han, Ping Luo, Gang Zeng, and Jingdong Wang.
\newblock Context {Autoencoder} for {Self}-supervised {Representation} {Learning}.
\newblock {\em International Journal of Computer Vision}, 132(1):208--223, January 2024.

\bibitem{tong_videomae_2022-1}
Zhan Tong, Yibing Song, Jue Wang, and Limin Wang.
\newblock {VideoMAE}: {Masked} {Autoencoders} are {Data}-{Efficient} {Learners} for {Self}-{Supervised} {Video} {Pre}-{Training}, October 2022.
\newblock arXiv:2203.12602 [cs].

\bibitem{feichtenhofer_masked_2022}
Christoph Feichtenhofer, Haoqi Fan, Yanghao Li, and Kaiming He.
\newblock Masked {Autoencoders} {As} {Spatiotemporal} {Learners}, October 2022.
\newblock arXiv:2205.09113 [cs].

\bibitem{huang_masked_2023}
Po-Yao Huang, Hu~Xu, Juncheng Li, Alexei Baevski, Michael Auli, Wojciech Galuba, Florian Metze, and Christoph Feichtenhofer.
\newblock Masked {Autoencoders} that {Listen}, January 2023.
\newblock arXiv:2207.06405 [cs, eess].

\bibitem{devlin_bert_2019}
Jacob Devlin, Ming-Wei Chang, Kenton Lee, and Kristina Toutanova.
\newblock {BERT}: {Pre}-training of {Deep} {Bidirectional} {Transformers} for {Language} {Understanding}, May 2019.
\newblock arXiv:1810.04805 [cs].

\bibitem{bushuiev_emergence_2024}
Roman Bushuiev, Anton Bushuiev, Raman Samusevich, Corinna Brungs, Josef Sivic, and Tomáš Pluskal.
\newblock Emergence of molecular structures from repository-scale self-supervised learning on tandem mass spectra, April 2024.

\bibitem{jensen_identification_2024}
Mathias~N. Jensen, Eduarda~M. Guerreiro, Agustin Enciso-Martinez, Sergei~G. Kruglik, Cees Otto, Omri Snir, Benjamin Ricaud, and Olav~Gaute Hellesø.
\newblock Identification of extracellular vesicles from their {Raman} spectra via self-supervised learning.
\newblock {\em Scientific Reports}, 14(1):6791, March 2024.

\bibitem{bjerrum_data_2017}
Esben~Jannik Bjerrum, Mads Glahder, and Thomas Skov.
\newblock Data {Augmentation} of {Spectral} {Data} for {Convolutional} {Neural} {Network} ({CNN}) {Based} {Deep} {Chemometrics}, October 2017.
\newblock arXiv:1710.01927 [cs].

\bibitem{zhang_genotype_2024}
Yirui Zhang, Kai Chang, Babatunde Ogunlade, Liam Herndon, Loza~F Tadesse, Amanda~R Kirane, and Jennifer~A Dionne.
\newblock From {Genotype} to {Phenotype}: {Raman} {Spectroscopy} and {Machine} {Learning} for {Label}-{Free} {Single}-{Cell} {Analysis}.
\newblock {\em ACS Nano}, 2024.

\bibitem{ho_rapid_2019}
Chi-Sing Ho, Neal Jean, Catherine~A. Hogan, Lena Blackmon, Stefanie~S. Jeffrey, Mark Holodniy, Niaz Banaei, Amr A.~E. Saleh, Stefano Ermon, and Jennifer Dionne.
\newblock Rapid identification of pathogenic bacteria using {Raman} spectroscopy and deep learning.
\newblock {\em Nature Communications}, 10(1):4927, October 2019.
\newblock Number: 1 Publisher: Nature Publishing Group.

\bibitem{horgan_high-throughput_2021}
Conor~C. Horgan, Magnus Jensen, Anika Nagelkerke, Jean-Philippe St-Pierre, Tom Vercauteren, Molly~M. Stevens, and Mads~S. Bergholt.
\newblock High-{Throughput} {Molecular} {Imaging} via {Deep}-{Learning}-{Enabled} {Raman} {Spectroscopy}.
\newblock {\em Analytical Chemistry}, 93(48):15850--15860, December 2021.

\bibitem{vaswani_attention_2023}
Ashish Vaswani, Noam Shazeer, Niki Parmar, Jakob Uszkoreit, Llion Jones, Aidan~N. Gomez, Lukasz Kaiser, and Illia Polosukhin.
\newblock Attention {Is} {All} {You} {Need}, August 2023.
\newblock arXiv:1706.03762 [cs].

\bibitem{chen_simple_2020}
Ting Chen, Simon Kornblith, Mohammad Norouzi, and Geoffrey Hinton.
\newblock A {Simple} {Framework} for {Contrastive} {Learning} of {Visual} {Representations}, June 2020.
\newblock arXiv:2002.05709 [cs, stat].

\bibitem{li_contrastive_2020}
Yunfan Li, Peng Hu, Zitao Liu, Dezhong Peng, Joey~Tianyi Zhou, and Xi~Peng.
\newblock Contrastive {Clustering}, September 2020.
\newblock arXiv:2009.09687 [cs, stat].

\bibitem{eldele_time-series_2021}
Emadeldeen Eldele, Mohamed Ragab, Zhenghua Chen, Min Wu, Chee~Keong Kwoh, Xiaoli Li, and Cuntai Guan.
\newblock Time-{Series} {Representation} {Learning} via {Temporal} and {Contextual} {Contrasting}.
\newblock In {\em Proceedings of the {Thirtieth} {International} {Joint} {Conference} on {Artificial} {Intelligence}}, pages 2352--2359, Montreal, Canada, August 2021. International Joint Conferences on Artificial Intelligence Organization.

\bibitem{zhang_supporting_2021}
Dejiao Zhang, Feng Nan, Xiaokai Wei, Shang-Wen Li, Henghui Zhu, Kathleen McKeown, Ramesh Nallapati, Andrew~O. Arnold, and Bing Xiang.
\newblock Supporting {Clustering} with {Contrastive} {Learning}.
\newblock In {\em Proceedings of the 2021 {Conference} of the {North} {American} {Chapter} of the {Association} for {Computational} {Linguistics}: {Human} {Language} {Technologies}}, pages 5419--5430, Online, 2021. Association for Computational Linguistics.

\bibitem{selvaraju_grad-cam_2017}
Ramprasaath~R. Selvaraju, Michael Cogswell, Abhishek Das, Ramakrishna Vedantam, Devi Parikh, and Dhruv Batra.
\newblock Grad-{CAM}: {Visual} {Explanations} from {Deep} {Networks} via {Gradient}-{Based} {Localization}.
\newblock In {\em 2017 {IEEE} {International} {Conference} on {Computer} {Vision} ({ICCV})}, pages 618--626, Venice, October 2017. IEEE.

\end{thebibliography}

\end{document}